# Near-inertial wave propagation between stratified and homogeneous layers

**by Hans van Haren**

NIOZ Royal Netherlands Institute for Sea Research, P.O. Box 59, 1790 AB  Den Burg, the Netherlands
E-mail: hans.van.haren@nioz.nl

**Abstract**

The propagation of inertio-gravity waves (IGW) into the deep-sea is relevant for energy transfer to turbulence where waves break, and thus for redistribution of nutrients, oxygen and suspended matter. In constant stratification, vertical IGW-propagation is readily modelled. In varying stratification, where homogeneous layers alternate with stratified layers, transmission and reflection cause complex patterns. Half-year long moored acoustic Doppler current profiler (ADCP) observations midway between the Balearic Islands and Sardinia in the 2800-m deep Western-Mediterranean Sea occasionally demonstrate a distinct transition, between weakly stratified ($N \geq 2f$) and homogeneous ($N \leq f$) layers, of IGW at near-inertial frequencies. Here, N denotes the buoyancy frequency and f the local inertial frequency (vertical Coriolis parameter). The transition in stratification is rather abrupt, within $\Delta z = 25$ m and provides an amplitude-reduction of 1.3 for super-inertial motions. Simulations with non-traditional momentum equations involving the horizontal Coriolis parameter $f_h$ qualitatively confirm observed IGW-refraction. The observational area is marked by variations in hydrographic characteristics, with abundant mesoscale eddies to the south and dense-water formation to the north of the site during the previous winter. Thus, also transitions occur from deep homogeneous layers into deeper, recently formed stratified ones. Polarization spectra of shear are bound by IGW-limits related to N=f, while current-polarization to $2f_h$. These frequencies coincide with large-scale buoyancy frequencies independently observed in various layers using shipborne CTD-profiling.

# 1 Introduction

Classically at all wave numbers, freely propagating inertio-gravity waves (IGWs) have frequencies smaller than the buoyancy frequency N and greater than the inertial frequency (vertical Coriolis parameter) $f = 2\Omega\sin\varphi$, $\Omega$ denoting the earth rotational vector-amplitude and $\varphi$ the latitude at the site of observations. Motions at near-inertial frequencies follow from geostrophic adjustment, e.g. associated with the passage of atmospheric disturbances or frontal collapse (Gill 1982). A tropical cyclone for example, may cause immediate near-inertial response at depths of around 1000 m (Shay and Elsberry 1987) and even deeper than 4000 m (Morozov and Velarde, 2008) due to ocean-level variations. Such response is followed by downward propagating near-inertial waves at speeds the increase up to about 0.01 m s$^{-1}$ through weaker stratification at great depths (Morozov and Velarde 2008). Of concern is the possibility of deep penetration, with little amplitude-attenuation, of IGWs at f in the ocean interior while crossing layers of varying stratification including (weakly) stratified ones, in which $N \geq 2f$, and homogeneous ones more than 100-m thick, in which $N \leq f$. A better understanding of, in particular near-inertial IGW-energy penetration, hence of the displacement of shear and convection induced by breaking IGW, is relevant for deep-ocean turbulent mixing.

Under the traditional approximation, momentum-equation terms involving the horizontal component of the earth rotation (the horizontal Coriolis parameter) $f_h = 2\Omega\cos\varphi$ are neglected. As a result of this model-approximation, any transition from a stratified layer to a homogeneous one would completely reflect propagating IGWs (Munk 1980), while only exponentially decaying amplitudes are predicted in the homogeneous layer. Although the importance of $f_h$ for the generation of IGWs in weakly stratified and notably homogeneous layers has been known for half a century or more (LeBlond and Mysak 1978; review by Gerkema et al. 2008), it has been largely ignored within the oceanographic community. Neglecting $f_h$ is motivated by the idea that the upper ocean over large vertical scales O(100-1000 m) is, in general, strongly stratified in density. However, the ocean is certainly not strongly stratified in the deep below 2000 m and in areas where, also close to the surface, double-diffusion or convective dense-water formation are dominant. Homogeneous convective layers that can have considerable vertical thickness O(100 m) have been observed in areas like the Arctic (Timmermans et al. 2007), around the equator (Dengler and Quadfasel 2002) and at mid-latitudes, e.g. in two basins with similar oceanographic dynamics: in the Sea of Japan (Talley et al. 2003; Matsuno et al. 2015) and in the Western-Mediterranean Sea (van Haren and Millot 2004).

For the latter area, two observational studies in the southern Algerian sub-basin (van Haren and Millot 2004; 2005) suggested near-inertial IGW-propagation, following un-attenuated passage, across the interface between stratified and homogeneous layers. Near-inertial IGW-motions in homogeneous layers generally have marked rectilinear polarization in the horizontal plane [u, v] and show a large aspect ratio of elevated vertical (over horizontal) velocities $w(f)/[u, v](f) = 0.1$-$1$. In stratified layers of the ocean, near-inertial horizontal polarization is much more circular and aspect ratios are only O(0.01-0.001).

In convective chimneys of a dense-water formation area, vertical currents are easily distinguished from those at near-inertial frequencies (Voorhis and Webb 1970; Gascard 1973; Schott and Leaman 1991). The former vary more rapidly, O(1 hour) or less, in both frequency and amplitude, and they are much more irregular when registered at a mooring, compared to motions at a particular frequency like f. Although vertical currents associated with deep-water formation have been extensively studied (see review by Marshall and Schott (1999), the importance of relatively large $w(f)$ is not explored yet to its full extent.

In this paper, the focus is on $w(f)$ related to near-inertial IGW-propagation, and on homogeneous layers that are much thicker than the typical buoyancy scale O(10-100) m (Fig. 1b). Specifically, the $w(f)$-transition is studied between stratified and homogeneous layers using observations and a simulation. Also, the relationship is considered between observed spectral IGW characteristics and stratification.

The observational site is located in the Western-Mediterranean Sea, roughly halfway between areas to the north where dense waters are formed at times and areas to the south where mesoscale eddy activity is intense (Fig. 1a). Occasionally, newly formed dense-water masses are seen passing, with vigorous motions reaching speeds easily up to 0.5 m s$^{-1}$ near the ~2000-m deep seafloor (Millot and Monaco 1984), which value is more than three times the 'common' values. These motions are alternated with motions associated with the northern extent of anticyclonic eddies originating from the Algerian slope that can create some kind of large-scale eastward flow at the mooring site. Both dense-water formation and eddies may induce vigorous vertical motions, with similar amplitudes but on difference scales compared to w(f). Near-inertial convective mixing does not necessarily occur near the sea-surface but rather in the ocean interior, that is in layers capped above and below by well-stratified layers (van Haren and Millot 2005). It will be shown using simulation and observations that, for both the wave propagation and the suggested mixing, consideration of $f_h$ is crucial.

**2 Data**

A single mooring was deployed at 40°0′ N, 6°0′ E (H = 2760 m water depth) between 13/04/2005 and 12/02/2006 (Fig. 1a). The mooring consisted of 6 single-point current meters and one 75-kHz, four-beam acoustic Doppler current profiler (ADCP) mounted in an elliptical buoy with an upper main buoyancy element at z = -1630 m. Mooring characteristics are given in Table 1. Except one, all current meters delivered good data throughout the deployment period. Unfortunately, the ADCP stopped, due to programming errors, halfway the intended period on 20/09/2005, two days before a major change in observations, see Section 3. Its data were of reasonable quality over a range of about 300 m and noisy above that due to a lack of sufficient acoustic scatterers. All four current meters deployed above the ADCP were seen in its beams, but clearly distinguishable from unbiased data.

The ADCP's w-data quality is verified using the additional 'error velocity' e parameter, defined as the difference between the two w-estimates computed from the two beam pairs. Heterogeneities in u, v, w between the ADCP-beams, for example caused by an object obstructing one or more beams, will affect w-measurements. This is noticeable in e diverting from white noise and approaching the level of w. In the frequency range of interest, $|w| > 10|e|$ is observed at z = -2050 m (see Section 3) and we conclude that w(f) is significantly measured. As a result, good quality daily-averaged w are expected to a relative accuracy of about $3 \cdot 10^{-4}$ m s$^{-1}$ (Table 1, in which error levels for 1200-s sampling periods are given), which is at least one order of magnitude smaller than typically observed inertial amplitudes (van Haren and Millot 2005).

The four acoustic current meters also provided estimates of w, but these data are much noisier than the ADCP-w (Table 1). Therefore, most important parameters from these instruments are current components u and v, which are used for specifying the horizontal current polarization, and temperature T, which is used to monitor the variations in water mass properties with time, reflecting passages of eddies and newly formed dense water. Compared to the single good working mechanical current meter the acoustic devices showed larger amplitudes by 20%, as was inferred using the weak tidal amplitudes as measure. The mechanical current meter amplitude was corrected accordingly.

During the deployment and recovery cruises shipborne SeaBird-911 Conductivity-Temperature-Depth (CTD) profiles were obtained. These data are used for independent information on stratification variations in the vertical. Due to cable-limitations, the 2005-profile was stopped at z = -2500 m.

## 3 Observations

### 3.1 Background data

The density profiles observed using CTD in the deep (Fig. 1b) show a weak stratification that is mainly characterized by a homogeneous layer in between two weakly stratified ones. Typical values of the buoyancy frequency at some nominal depths are: $N(z = -1500\text{ m}) \equiv N_{1500} = 2f_h$, $N_{1800} = 0$ (which is difficult to distinguish from $N = f$), $N_{2400} = 4f_h$, $N_{2700} = 6f_h$. The former three values can be explained in terms of stability, gradient Richardson number, and analysis considering 'slantwise convection' in the direction of $\Omega$ (Straneo et al. 2002; van Haren 2008). The density profiles show abrupt variations between the different layers, occurring within $O(10\text{ m})$, but they do not vary strongly with time. However, during the first half of the mooring period, T and salinity S were very heterogeneous, as observed in yoyo-CTDs (van Haren and Millot 2009) and from T-time-series (Fig. 2b). Between the CTD-observations in April 2005 and February 2006, the homogeneous layer density decreased by a small but significant amount (Fig. 1b), which is equivalent to an adiabatic T-increase of 0.013°C, on average. This value is smaller by half an order of magnitude compared to the T-variations recorded with time at any particular depth (e.g., Fig. 2b). As a result, the latter variations are at least in half due to other, local mixing processes and not due to remote dense-water formation.

The temperature variations with time (Fig. 2b) are accompanied by relatively large variations in current speeds (Fig. 2a), with typical values of $0.2\text{ m s}^{-1}$ before day 265, and less than half that value during the 'low-current' period after day 265. Peak values reach up to $0.45\text{ m s}^{-1}$, which seriously deflect the mooring that was designed for the lower water-flow speeds above. The peak period of day 109 is not considered here, as the approximately 20°-tilt (Fig. 2c) caused the ADCP-compass to shut down. During occasional 10°-tilts, the instruments were vertically lowered by about 10 m, or one vertical bin size for the ADCP.

Nevertheless, the time-series between days 110 and 265 during spring and summer 2005 showed some 10-day periods of relatively constant temperature and current speeds, e.g. between days 140-150. During such periods, $|w| \gg |e|$ near the ADCP (Fig. 2d), not only in low-frequency variations that are negatively biased due to the passage of meso-scale eddies (van Haren et al. 2006), but also in relatively higher frequency variations that reflect near-inertial motions, not instrumental noise, as will be demonstrated in the next sub-section.

**3.2 Spectra**

The two periods, before and after day 265, show distinctly different kinetic energy spectra (Fig. 3; spectra were computed over days 125-260 and 270-405 to have periods of similar lengths). During the first period compared to the second, energy levels are higher at nearly all frequencies and depths except for near-inertial motions below z < -2350 m (Fig. 3a,b; green and purple spectra). Near-inertial peaks at all depths (except -2720 m, purple spectrum) are much more similar during the first period (Fig. 3a) than during the second one (Fig. 3b). This indicates some relatively simple downward continuous propagation during the first period as compared to a less simple propagation during the second one.

High-frequency (noise) energy for frequencies $\sigma > N_{1500}$ is larger by a factor of about two during the first period. A notable much better spectral resolution is found for the mechanical current meter compared to the acoustic ones, which is intrinsic (cf. Table 1) and not due to being located at different depths. However, especially the sub-inertial motions $\sigma \sim< 0.8f$, comprising meso-scale activity, are most enhanced by up to two orders of magnitude during the first period compared to the second. This has major effect on the inertial motions, at z = -2720 m (purple), and, especially -2350 m (green) where they are more pronounced than the others during the second, low-current period (Fig. 3b). During this period the f-motions are more circular (Fig. 3d) than during the first period (Fig. 3c). This is also observed for vertical current differences (shear) (Fig. 3e,f) and at shallower depths, where inertial amplitudes are somewhat smaller by factors up to 1.5 compared to the first period. Especially during the second period, horizontal current polarization is distributed less symmetrically around f than shear. It seems that current polarization is bound by IGW-non-traditional limits [$\sigma_{min}$, $\sigma_{max}$] (for N=4$f_h$ related to z = -2400 m; blue-dashed lines) whereas shear is bound by [$\sigma_{min}$, $\sigma_{max}$] (for N=f). Here, the stratification is inferred from the spectral limits, using IGW-frequency limits defined by (Saint-Guily 1970; LeBlond and Mysak 1978; Gerkema and Shrira 2005):

$$\sigma_{min}^2 = s - (s^2 - f^2 N^2)^{1/2} < f^2, \quad (1a)$$
$$\sigma_{max}^2 = s + (s^2 - f^2 N^2)^{1/2} > N^2, \quad (1b)$$

where $2s = N^2 + f^2 + f_s^2$,
and $f_s = f_h \sin\alpha$, $\alpha$ the angle in the horizontal plane with respect to east, including $f_h$,

which contributes in meridional direction.

Three remarkable observations are noted in the present data. First, a peak is observed at $\sigma$ = $f_h$, albeit only in near-bottom current data during the second period (purple graph in Fig. 3b). This suggests a direct observation of $f_h$-dynamics driving u,w in weakly stratified layers in the same manner as u,v-inertial motions in strong stratification. The $f_h$-peak is found at a depth where one expects sub-inertial IGWs to be trapped, not gyroscopic waves (Gerkema and Shrira 2005; van Haren 2006). Such trapping is predicted for just sub-inertial IGWs in weak stratification, f < N ~< 5f, when they propagate poleward in an environment of larger inertial frequency at higher latitude, and capped by larger stratification above them.

Second, polarization increases, becoming more circular, with depth, from blue (z = -1785 m) to green (-2350 m) in both periods (Fig. 3c,d). With increasing polarization, the near-inertial peak-frequency varies from super- to sub-inertial. Both observations confirm the trapping of sub-inertial motions in (deep) stratified layers and super-inertial motions in (intermediate) homogeneous layers.

Third, especially during the second period (Fig. 3b) in the near-bottom (purple) kinetic energy spectrum, one observes elevated energy levels of high-frequency IGW that significantly start to roll-off at $\sigma \approx 2f_h$ (hydrographically observed $N_{1500}$), before finally resuming the general slope at $\sigma \approx 4f_h$ (observed $N_{2400}$). Apparently, motions are driven under support of stratification on both sides of the N=0 layer, possibly during different periods for each stratified layer.

A similar spectral roll-off in the range $2f_h < \sigma < 4f_h$ is also observed in w-spectra higher-up, measured by ADCP (Fig. 4). At all frequencies $\sigma \leq 2f_h$, w-variance extends above e-variance by half an order of magnitude on average, which implies that w is measured significantly. The w-roll-off starts at $\sigma \approx 2f_h$, reaches e (white noise level) at $\sigma \approx 4f_h$ and definitely is indistinguishable from e for $\sigma > 6f_h$ (equal to observed $N_{2700}$).

The w-spectrum (Fig. 4) does not show a specific peak at f, like in the horizontal kinetic energy spectrum, but a broad weakly elevated band between [0.57f, 1.76f]. These bounds correspond with [$\sigma_{min}$, $\sigma_{max}$] (for N=f) and more or less corresponding with the band of non-rectilinear shear (Fig. 3e). The largest, non-significant, w-"peak" is found at $\sigma = f_h$ (Fig. 4), which suggests non-traditional dynamics from a second observable at a different depth during a different period than above. It is not understood why no corresponding peak is observed in the horizontal kinetic energy spectrum at the same depth.

The observed smooth continuation of the w-spectrum from IGW to sub-inertial frequencies may point at an unknown and coincidental coupling between meso-scale and IGW motions. It may however also point at the smooth transition between IGW in density-stratified and homogeneous layers, as the latter are found in the range [0, 2$\Omega$] for meridional motions ($\alpha = \pi/2$ in (1)) and because the observations last O(1-10 days) when N = 0 locally. This may cause variations in time of both the inertial horizontal current polarization, which are no longer circular in homogeneous layers at mid-latitudes (van Haren and Millot 2004), and the vertical inertial currents, which are non-negligible in such layers (van Haren and Millot 2005). This will be pursued in the next subsection.

**3.3 Band-pass filtered time-series**

Evidence of large w([0, 2Ω]), including w(f), is not easy to find in observations, because locally measured w can be generated at remote density interfaces that determine their frequency. Nevertheless, if we use T as a tracer, and especially its adiabatic value as evidence for occurrence of N = 0, a comparison between T and band-pass filtered w shows some correspondence (Fig. 5). At f, elevated w-amplitudes that are uniform through the ADCP-range are found more during periods when T is low, at its mean adiabatic value. Mutually exclusive with w(f) are w($f_h$), which are less uniform across the ADCP-range (considering only the less-noisy lower 200 m). The w($f_h$) show smallest amplitudes during periods when w(f) are elevated, like 2$f_h$-motions (Fig. 5) that seem directly coupled with $f_h$-motions and which show a small spectral sub-peak (Fig. 4).

The observation that $f_h$- and 2$f_h$-motions are mostly found when the water column is stratified, i.e. temperature being elevated above local homogeneous-adiabatic values, is understandable for 2$f_h$-, but not for $f_h$-motions, unless the latter are sub-harmonics of the former. If the near-inertial motions, mainly found when N = 0, are super-f they are likely trapped motions; if they are sub-f they represent propagating gyroscopic waves (van Haren, 2006). To distinguish the latter from the former is complex, because the non-traditional approach for each motion provides two differently sloping characteristics (Gerkema and Shrira, 2005).

Focusing on most energetic and penetrable near-inertial motions, smooth transitions between N > f and N = 0 are monitored in some detail by considering all current components. Variations are large in depth and time. In a detailed example over an arbitrary 7-day period (Fig. 6) we observe relatively large w ≈ 0.3[u, v], with near-homogeneous values across most of the entire vertical range. Shear across independent vertical scales of Δz = 25 m is low below z < -1900 m, but reaches values of about 4$f_h$ at shallower z > -1900 m. It is noted that above z > -1750 m, e becomes relatively large and data should be ignored.

The phase speed of a parameter is inversely proportional to its amplitude slope in the vertical-time plane. Considering w(f), several changes in [z, t]-slope are observed, e.g. at day 181 near z = -1750 and near -2000 m, at day 185 near -2025 m (circles in Fig. 6). These changes are all abrupt in z, from one independent vertical position to the other (about 25 m apart). The slant-sloping amplitudes are less by a factor of 1.0-1.5 than the vertically uniform w(f). Such changes may point at a transition of super-inertial motions from stratified to homogeneous and homogeneous to (stronger) stratified layers, respectively. This is qualitatively verified with results from a simulation in Section 4.

## 4 Simulating transition

Adopting the non-traditional approach including the terms with $f_h$, near-inertial IGWs are special. This is because for arbitrary horizontal propagation direction $\alpha \neq 0$, only waves in the frequency band $(1-\varepsilon)f < \sigma < (1+\varepsilon)f$, $\varepsilon \ll 1$, can pass between stratified layers where buoyancy frequency $N > f$ and homogeneous layers where $N = 0$. The maximum extension of the pass-band will be for IGWs propagating in meridional, north-south direction, for which $f < \sigma < (f^2+f_s^2)^{1/2} = 2\Omega$ (e.g., Gerkema and Exarchou 2008). However, IGW-passages from layers with $N > f$ to $N = 0$ involve a subtle transition from classic propagating waves to trapped waves when maintaining a fixed frequency-relation with respect to f, being either sub- or super-inertial (Gerkema and Shrira 2005). They remain propagating waves only when they switch from sub- to super-inertial frequency, or vice versa, that is when propagating partially in meridional direction and under the specific condition that they propagate equatorward when crossing from homogeneous to stratified layers and poleward when crossing from stratified to homogeneous layers (van Haren 2006).

Following modelling by Gerkema and Exarchou (2008), an instantaneous solution of a linear 2-D z,y flat-bottom analytic simulation of IGW-propagation using 20 vertical modes in a non-traditional approach shows propagating waves from left to right, crossing an interface between homogeneous and uniformly stratified layers (Fig. 7). A single fixed frequency of motions is used in the calculation. Here, only results are shown for near-inertial frequency vertical motions, albeit distinction is studied between just-super- and just-sub-inertial motions. Only frequencies are chosen in the range (1) that will yield free IGW-propagation in both layers. This naturally limits the frequency-range to near-inertial frequencies.

For all subtle frequency variations, distinct changes are found in near-inertial characteristics-direction and -amplitude, by a factor of ~1.3, when waves cross smoothly a sharp abrupt change between stratified and homogeneous waters. Clearly, the asymmetric two solutions of characteristics (for $\alpha = \pi/2$) are observed per layer, with a vertical one in N = 0 having largest w-amplitude when $\sigma > f$. This corresponds with ADCP-observations in Fig. 6. Sub-inertial w-amplitudes are largest in stratified layers where they may partially trap.

In principle, the above simulation could compare well with some of the observations of abrupt phase changes in amplitude and direction of w across abrupt changes in density, provided we can transfer [z, t] to [z, y]. However, this is not so obvious (see also Gerkema and Exarchou (2008), despite some of the similar qualitative results given above, notably the abrupt changes in phase and associated amplitude variations.

## 5 Discussion

The presented simulation shows already complex propagation and trapping results that are hard to be distinguished in the observed time series. This lack of quantitative comparison is due to large spatio-temporal variations of wave amplitude and direction, even in a (Mediterranean or Japan) sea where tides are weak. Nonetheless, some similarities are found, which at least suggest that transitions of near-inertial motions across a stratified-homogeneous interface have been observed in ADCP-data and that such interface is abrupt ($\Delta z \leq 25$ m).

The abruptness of the interface is questioned by the smoothness of IGW-transition. IGWs in homogeneous layers, gyroscopic waves, have the earth's rotational vector governing their direction of propagation. Since horizontal motions are spectrally strictly bound by non-traditional IGW-bounds as defined from N = f, $2f_h$ or $4f_h$, it is suggested that the earth rotational axis sets the direction of free convection, which results in non-zero stratification in the direction of gravity: 'slantwise convection' (Straneo et al. 2002; Sheremet 2004).

As a result, inertial motions, the dominant propagating motions, may initiate similar convective mixing in stratified layers, which is equivalent to slantwise homogeneous, and vertically homogeneous layers, resulting in continuous T-S property variations across the stratification interface (e.g., van Haren and Millot (2009). By the same token, the difference in f-motions in the different layers will be differently polarized, and hence shear will change.

When temperature and salnity are not measured, an indication for local stratification is the computation of shear amplitude $|\mathbf{S}|$. When the stratification is marginally stable, to the point of overturning at gradient Richardson number Ri = $N^2/|\mathbf{S}|^2$ = 0.25-1.0 (Miles 1961; Howard 1961; Abarbanel et al. 1984), it can support shear up to $|\mathbf{S}| \leq 4f_h$. This has been observed in ADCP-data on 25-m scales (Fig. 6), and also on O(100 m) scales in current meter data (Fig. 8). In the deep between z = -2000 and -2350 m, maximum $|\mathbf{S}| = 2f_h$, which are mainly due to sub-inertial motions except around day 350 when inertial shear dominates even at this large vertical scale. Shallower between z = -1700 and -1800 m, maximum $|\mathbf{S}| = 4f_h$, whereas inertial shear is seldom larger than $|S(f)| = f \approx f_h$. These values are comparable to the local mean N-values, so that the stratification is, indeed, marginally stable. They also show that the water column is easily made locally convectively unstable during short periods of time.

To complicate things further, indications like variations in horizontal current polarization, changing from near-circular to increasingly elliptic, may vary strongly in the vertical, as has been shown in a numerical model (Gerkema and Exarchou, 2008). This may 'explain' the observed z,t-variable polarization for near-inertial horizontal motions (Fig. 9). Generally, near-f motions generate most shear because of their short vertical scales in stratified waters, especially in the Mediterranean where tides are also weak, resulting in generation of diapycnal turbulent overturning and mixing. Needless to say that the complex inertial internal wave propagation which is capable of un-attenuated border-crossing between homogeneous and stratified layers, will influence turbulent mixing in the deep. However, more future observational and modelling studies, also from other weakly stratified basins, are wanted for the precise mechanisms of deep-sea mixing that is so relevant for life in the abyss.

## 6 Conclusions

The deep Mediterranean observations show thick near-homogeneous layers in which horizontal current polarization is governed by non-traditional IGW-bounds related to $N = 2f_h$, and shear to $N = f$.

Non-traditional IGW-motions are also suggested from a sub-peak in kinetic energy at $f_h$ observed in a presumably stratified layer below a large homogeneous layer. It suggests partial trapping of poleward propagating IGW.

Vertical propagation of near-inertial IGW as monitored in the vertical current component varies rapidly with depth and time. This may be due to sub-mesoscale eddy activity in the area. The magnitude change by a factor of 1.3±0.2 is found in observations as well as in a simulation and confirms super-inertial motions vertically propagating from weakly stratified to homogeneous to deep stratified waters.

**Acknowledgments** I thank the crews of the R/V Thethys II and Le Suroît for the sea-operations, G. Rougier and C. Millot for preparing the 'GYROSCOP-2' mooring, and T. Gerkema for providing the model results. I gratefully acknowledge support from the Netherlands organisation for the advancement of scientific research, NWO, and Centre National de la Recherche Scientifique, CNRS, under the (alas no longer existing) French-Dutch scientific collaboration.

**Table 1.** GYROSCOP-2 mooring details. Aq = 2-MHz Nortek AquaDopp acoustic current meter, AR8 = Aanderaa RCM-8 mechanical current meter, ADCP = upward looking 75-kHz 20°-beam angle TeleDyne/RDI-LongRanger acoustic Doppler current profiler. hab = Height above bottom, I = echo intensity (acoustic amplitude), ACM = accuracy current measurement per sampling interval (si).

| Instrument | z (hab) [m] | si. [s] | Sensors | ACM $[10^{-2} m\ s^{-1}/si]$ |
|---|---|---|---|---|
| Aq | -1685 (1075) | 600 | u,v,w,p,T,I | u,v: 0.8; w: 1.2 |
| Aq | -1785 (975) | 600 | u,v,w,p,T,I | u,v: 0.8; w: 1.2 |
| AR8[+] | -1890 (870) | 1200 | T | -- |
| Aq | -1990 (770) | 600 | u,v,w,p,T,I | u,v: 0.8; w: 1.2 |
| ADCP | -1585/-2057 (703-1095) | 1200 (60 x 8 m) | 60(u,v,w,p,T,I) | u,v: 1.0; w: 0.3 |
| Aq | -2350 (410) | 600 | u,v,w,p,T,I | u,v: 0.8; w: 1.2 |
| AR8 | -2720 (40) | 1200 | u,v,T | u,v: 0.3 |

[+]: Current magnitude failed

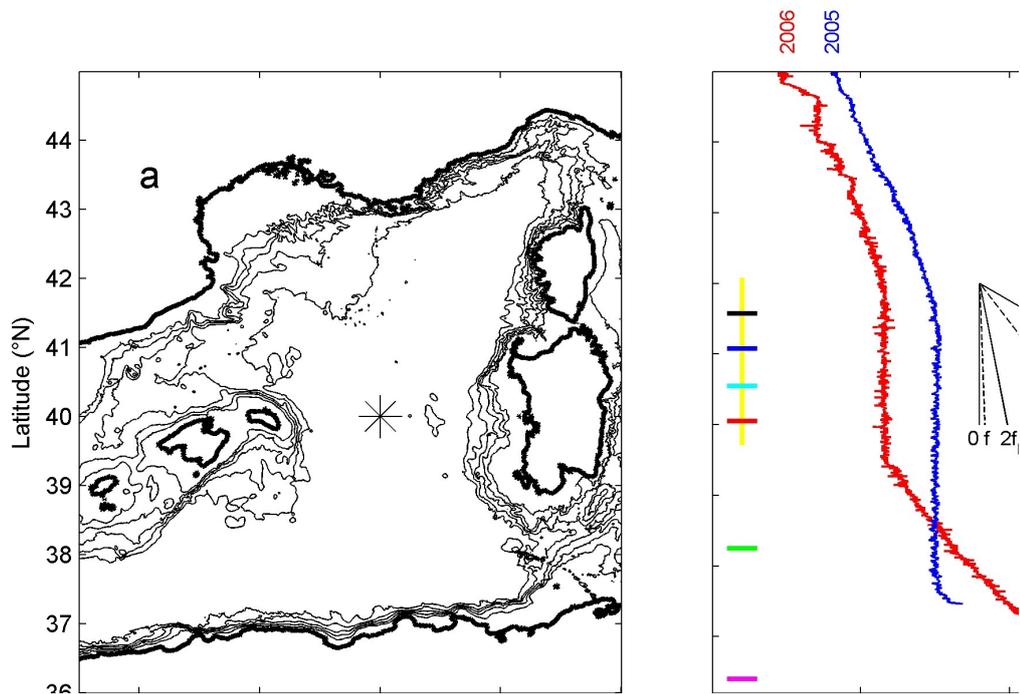

**Fig. 1** Site and hydrography. a) Western-Mediteranean Sea with observational site *. Depth contours every 500 m for [500, 2500] m and 2750 m. b) Typical local density anomaly profiles referenced to a pressure of 2000 dbar observed using shipborne CTD in April 2005 (blue; stopped at z = -2500 m) and February 2006 (red) during mooring deployment and recovery cruises, respectively. For reference, particular density gradient slopes are given (see text). To the left, the mooring is given schematically, with current meter (CM) depths (same colours as in Figs 2a,b, 3a,b; light-blue: only T-data) and range of upward looking ADCP (yellow). The local seafloor is at the x-axis.

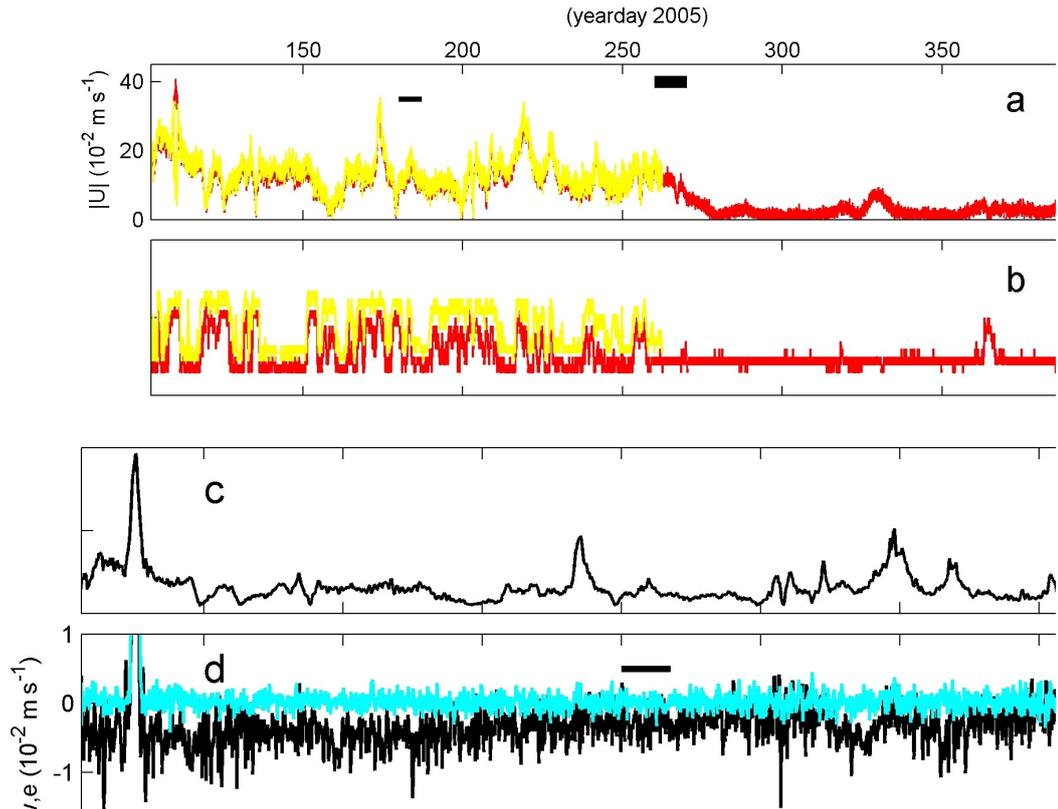

**Fig. 2** Time-series for the entire 10-month mooring period (a,b) and for the 5.5-month period of ADCP-data (c,d). Yeardays in 2006 are +365. a) Current amplitude at ADCP's bin 5 (z = -2025 m) and CM (-1990 m) using the instruments' colour coding in Fig. 1b. The horizontal bar indicates the period of Fig. 6. The ADCP record ends before a change in the time series' characteristics (days 260-270, indicated by thick black bar) that defines two different periods. b) T-time-series from the same instruments as in a), but at z = -2090 m for the ADCP. T-data are calibrated using CTD-profiles at recovery. During the 10-month period, the local homogeneous-adiabatic T raised by 0.013°C, which is approximately the T-sensors' resolution. c) Mooring-line tilt measured by the ADCP and reflecting water-flow drag. d) Vertical current-component (black) and error velocity (light-blue) measured at ADCP's bin 5 (z = -2025 m) and smoothed using a 3-h running mean. The horizontal black line indicates the period of Fig. 6.

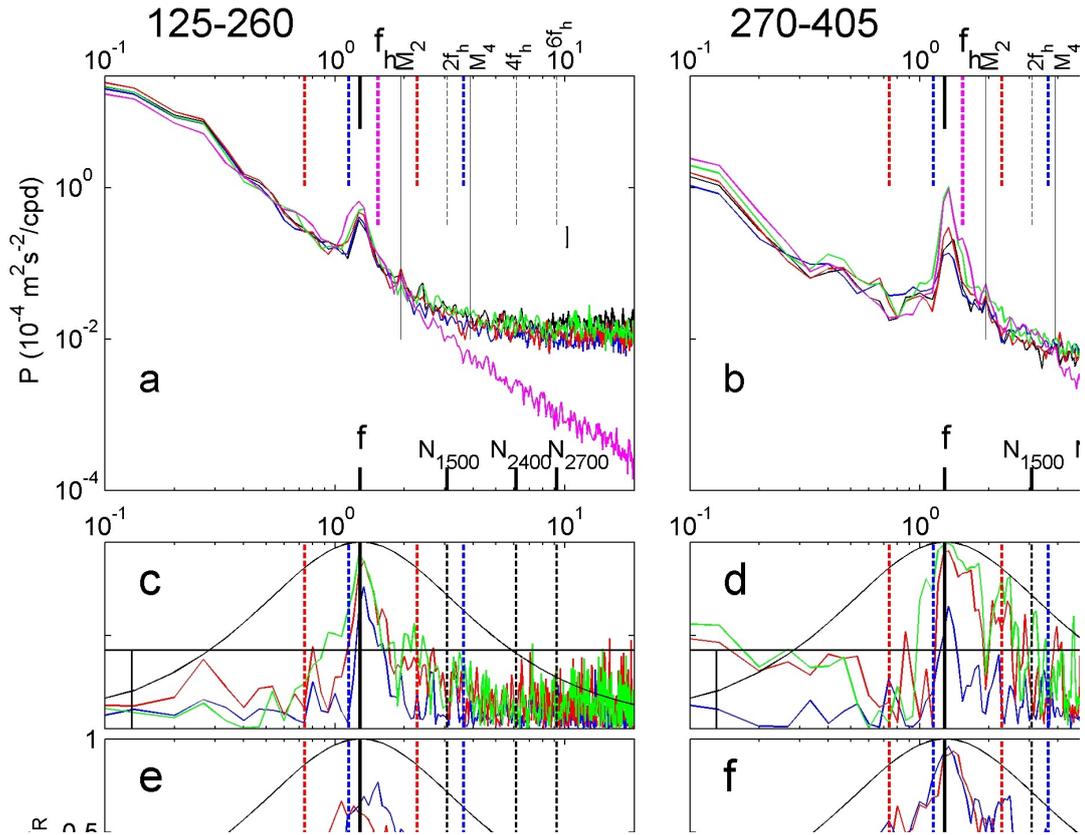

**Figure 3**. Kinetic energy (a,b) and rotary (c-f) spectra from horizontal (u, v)-data for two 135-day periods before (a., c., e.) and after (b., d., f.) the thick black bar in Fig. 2a. a,b) Kinetic energy spectra at each specific level with typical buoyancy frequencies (from Fig. 1b) such as $N_{1500} \approx 2f_h$, $N_{2400} \approx 4f_h$, and $N_{2700} \approx 6f_h$ indicated by thin-black dashed lines. The shorter red-dashed lines mark the inertio-gravity wave (IGW) limits computed using $N = f$ and the blue-dashed lines those using $N = 2f_h$. They may be compared with heavy black solid lines, indicating the traditional range [f, N], borders exclusive, where $N = 2, 4$ or $6f_h$. Thin solid lines indicate some specific tidal harmonic frequencies. Color-coding as in Fig. 1b, with obviously much less noise in the mechanical AR8 CM (purple). c,d) Rotary coefficient at z = -1785, -1990 and -2350 m. Values of 1 indicate circular motions, 0 indicates rectilinear motions. e,f) Rotary coefficient of current difference between z = -1685 and -1990 m (red), -1785 and -2350 m (blue). The smooth curve follows from linear theory (Gonella 1972), which formally only holds for $\sigma > f$ (N >> f); the horizontal line indicates 95% confidence value.

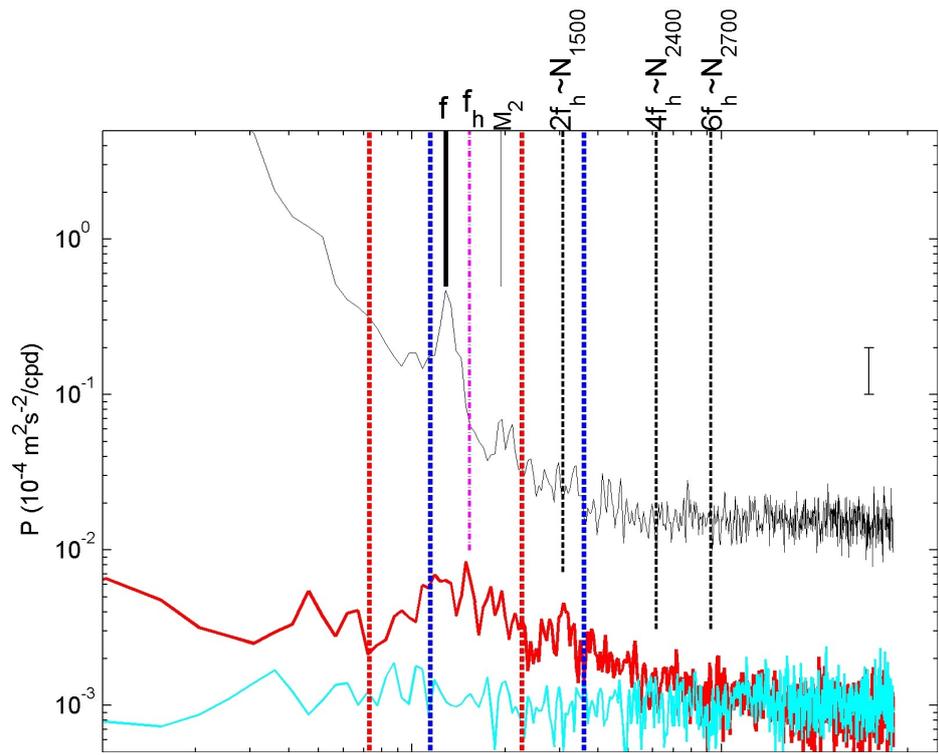

**Figure 4.** Spectra from ADCP-data at z = -2025 m. Horizontal kinetic energy (black), vertical current (red) and error velocity (light-blue). Vertical lines as in Fig. 3.

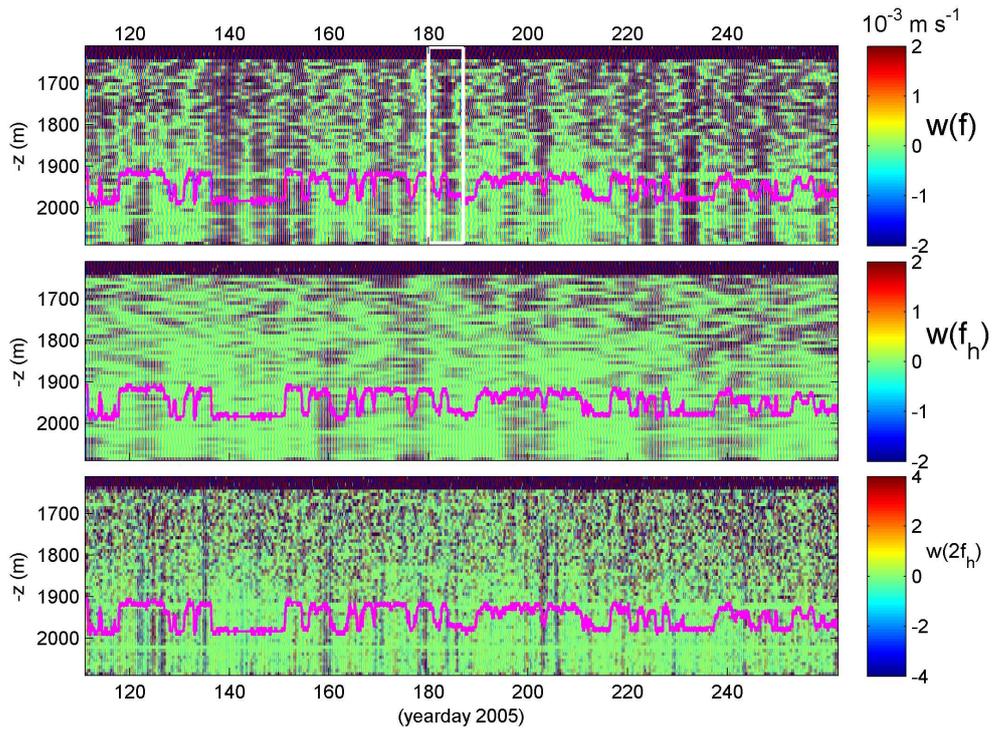

**Figure 5**. Band-pass filtered depth-time series of vertical currents. For reference, the time series of ADCP-temperature (yellow in Fig. 2b) is reproduced in purple (arbitrary scales). Note that data are noisy for z > -1900 m where patterns are less uniform. a) w at f (0.9f-1.1f band). The white box indicates the period of Fig. 6. b) w at $f_h$ (0.9$f_h$-1.1$f_h$ band). c) w at 2$f_h$ ($\sigma_{max}$(N=f)-$\sigma_{max}$(N=2$f_h$) band, i.e. right-side red- and blue-dashed lines in, e.g., Fig. 4). Note the different color scale compared to the upper panels.

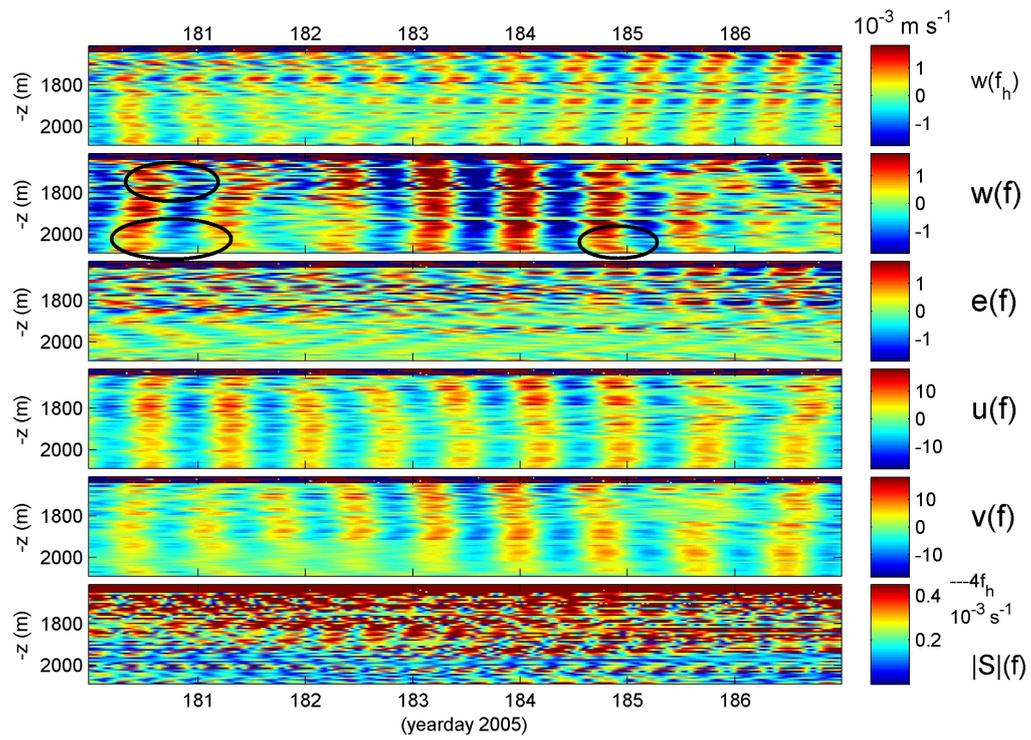

**Figure 6**. Detail of Fig. 5 for more parameters. All panels are band-pass filtered (0.9f-1.1f) inertial motions, except for the upper panel (0.9f$_h$-1.1f$_h$). Note the different color scales and units.

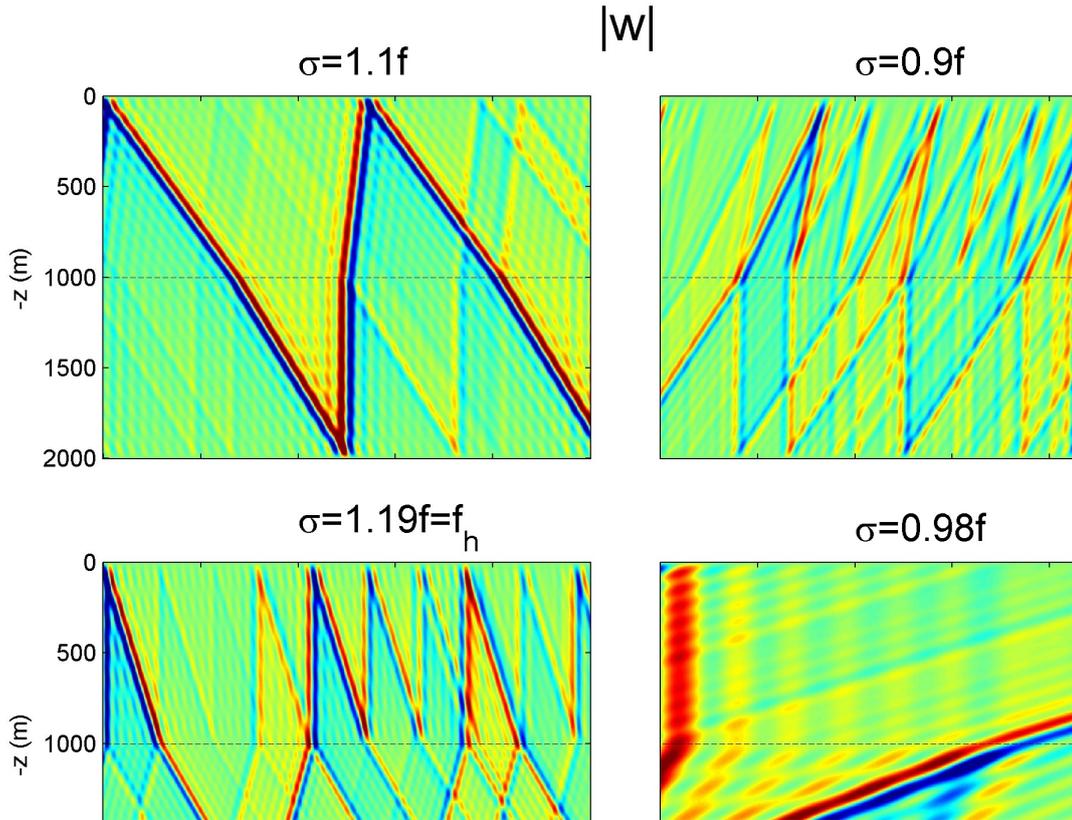

**Figure 7**. Simulation results of vertical current amplitudes |w| for internal wave 'beams' in a vertical-horizontal plane with the seafloor at z = -2000 m. The upper two panels show a transition (dashed line) between a weakly stratified layer above a homogeneous one. The lower two panels show a transition between a stratified layer below a homogeneous one. The left two panels are for super-inertial motions, the right two for sub-inertial motions. The left panels show amplitude enhancement and trapping in the homogeneous N=0 layer, while the right panels show enhancement and trapping in the stratified layer. Note the two different angles to the horizontal of up- and down-going rays, which is typical for the non-traditional approach (Gerkema et al. 2008).

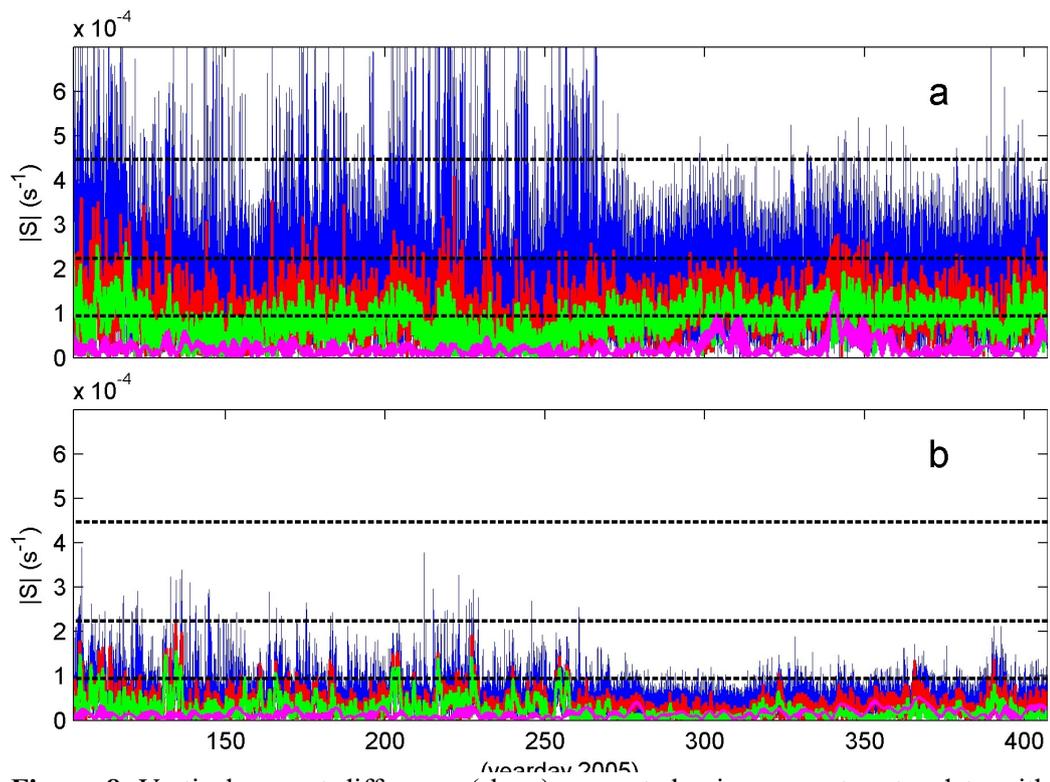

**Figure 8.** Vertical current difference (shear) computed using current meter data with large-scale 'Δz'. (a) Between z = -1685 and -1785 m. b) Between -1990 and -2350 m. Several filtered records are shown: total (in blue), low-pass filtered with cut-off at $4f_h$ (red), sub-inertial (green), and inertial (purple).

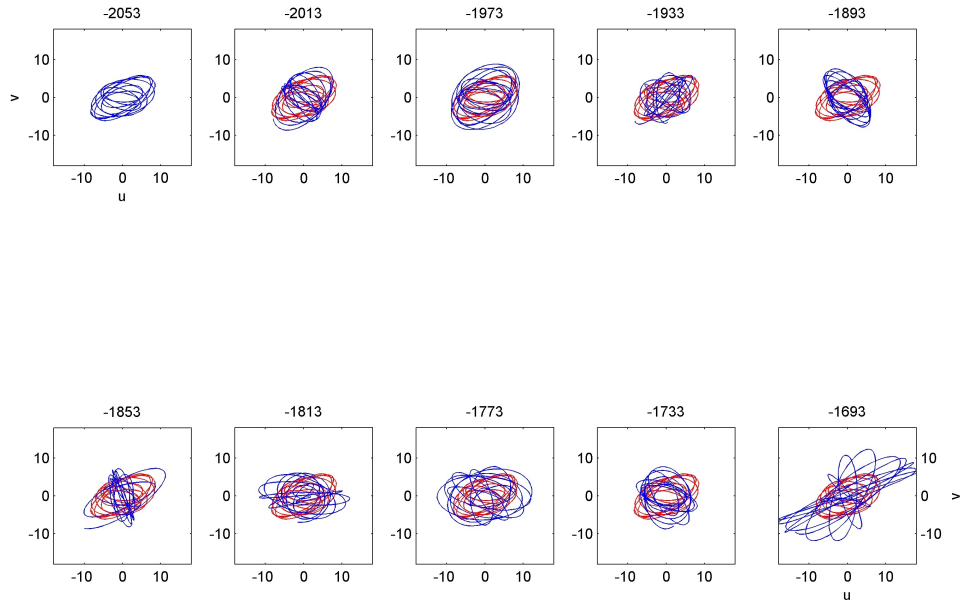

**Figure 9**. ADCP horizontal current hodographs in blue every 40 m for band-pass filtered inertial motions from the period in Fig. 6. The hodograph at z = -2053 m is plotted in red in all other panels for reference. Depths are given in m, horizontal currents in $10^{-3}$ m s$^{-1}$.